\documentclass[]{raa}
\usepackage{tipa}
\usepackage{bbm}
\usepackage{mathrsfs}
\usepackage{txfonts}
\usepackage{amssymb}            
\usepackage{graphicx,times}
\usepackage{natbib}

\providecommand{\bm}[1]{\mbox{\boldmath$#1$\unboldmath}}

\begin{document}

   \title{ A multi-transition molecular line study of infrared dark cloud
 G331.71+00.59
}

 \volnopage{ {\bf 20xx} Vol.\ {\bf 9} No. {\bf XX}, 000--000}
   \setcounter{page}{1}

   \author{Naiping Yu\inst{1,2}, Jun-Jie Wang\inst{1,2}
   }

   \institute{National Astronomical Observatories, Chinese Academy of Sciences,
             Beijing 100012, China; {\it yunaiping09@mails.gucas.ac.cn}\\
        \and
             NAOC-TU Joint Center for Astrophysics, Lhasa 850000, China
        \and
\vs \no
   {\small Received [year] [month] [day]; accepted [year] [month] [day]
} }

\abstract{Using archive data from the Millimeter Astronomy Legacy
Team Survey at 90 GHz (MALT90), carried out by Mopra 22-m telescope,
we made a first multi-transition molecular line study of infrared
dark cloud (IRDC) MSXDC G331.71+00.59. Two molecular cores were
found embedded in this IRDC. Each of these cores is associated with
a known ``extended green object" (EGO), indicating places of massive
star formation. The HCO$^{+}$ (1-0) and HNC (1-0) transitions show
prominent blue or red asymmetric structures, suggesting outflow and
inflow activities of young stellar objects (YSOs). Other detected
molecule lines include H$^{13}$CO$^{+}$ (1-0), C$_2$H (1-0), HC$_3$N
(10-9), HNCO(4$_{0,4}$-3$_{0,3}$), SiO (2-1) typical of hot cores
and outflows. We regard the two EGOs are evolving from IRDC to hot
cores. Using public GLIMPS data, we investigate the spectral energy
distribution of EGO G331.71+0.60, supporting this EGO is a massive
young stellar object (MYSO) driving outflow. G331.71+0.58 may be at
an earlier evolutionary stage.  }

   \authorrunning{Naiping Yu, Jun-Jie Wang.}            
   \titlerunning{A multi-transition molecular line study of infrared dark cloud
 G331.71+00.59}  
   \maketitle


%
%
\section{INTRODUCTIONS}           
Infrared dark clouds (IRDCs) were first discovered in the mid 1990s
by the two infrared satellites ISO (Perault et al. 1996) and MSX
(Egan et al. 1998) as silhouettes against the bright mid-infrared
Galactic background. Simon et al. (2006) identified 10931 candidate
IRDCs based on the MSX 8 $\mu$m data of the Galactic plane from
$\ell$ = 0-360$^{\circ}$ and $|$b$|$ $\leq $ 5$^{\circ}$. IRDCs are
predominantly found in the first and fourth Galactic quadrants and
near the Galactic mid-plane (Jackson et al. 2008), where the
mid-infrared background is greatest. Molecular line and dust
continuum studies of IRDCs have shown that they are cold ($<$25 k),
dense (n(H$_2$) $>$ 10$^5$ cm$^{-3}$, N(H$_2$) $>$ 10$^{22}$
cm$^{-2}$), and massive ($\sim$ 10$^2$ - 10$^5$ M$_{\odot}$)
structures with sizes of $\sim$ 1.15 pc. However, only small samples
of the IRDCs originally published by Simon et al. (2006) have been
investigated.

Cyganowski et al. (2008) identified more than 300 Galactic extended
4.5 $\mu$m sources, naming extended green objects or ``green
fuzzies'' for the common color coding of the 4.5 $\mu$m band as
green in Spitzer Infrared Array Camera (IRAC) three-color images.
They have suggested that the 4.5 $\mu$m IRAC band offers a promising
new approach for identifying MYSOs with outflows as it is supposed
to be due to H$_2$ ($\nu$ = 0-0, S (9, 10, 11)) lines and CO ($\nu$
= 1-0) band heads (Reach et al. 2006). The majority of EGOs are
associated with IRDCs. The infrared dark cloud MSXDC G331.71+00.59
(Fig.1) has almost never been studied even after it was found to be
associated with two EGOs: G331.71+0.58 $\&$ G331.71+0.60 (hereafter
G0.58 $\&$ G0.60). We here present the first molecular line study of
this source. The distance of G331.71+00.59 is still unknown.
According to the Galactic rotation model of Fich et al. (1989) (with
R$_{\odot}$=8.2 kpc and $\upsilon$$_{\odot}$=220 km s$^{-1}$ ), we
obtain a kinematic distance of either 4.26 kpc or 10.70 kpc. This
ambiguity arises because we are studying a region in the fourth
Galactic quadrant, where a given velocity may be associated with two
possible distances. As IRDCs are silhouettes against the Galactic
background, we use the near distance of 4.26 kpc in the following
analysis.

\begin{figure}
\onecolumn \centering
\includegraphics[height=2.7in,width=5.5in]{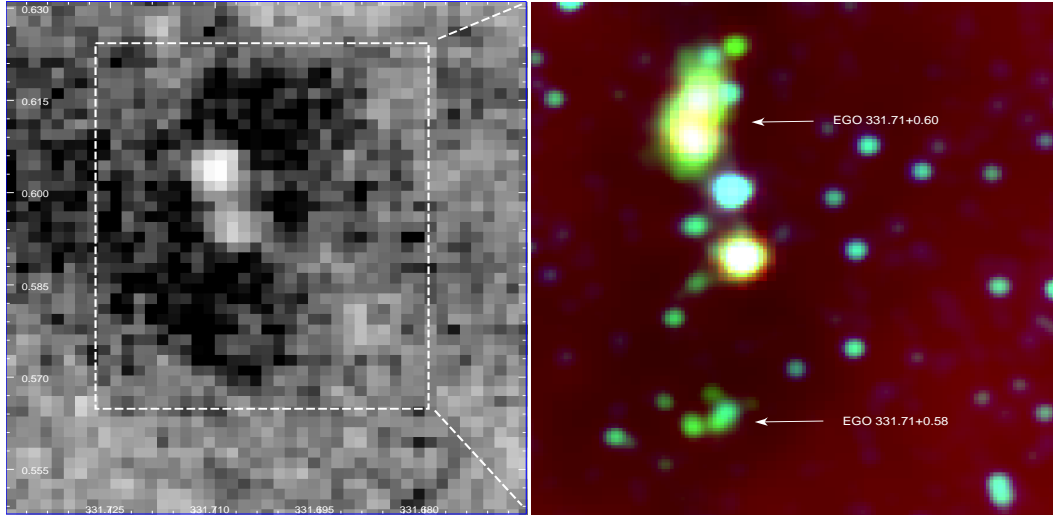}
\caption{Left: MSX A band image of IRDC G331.71+00.59. Right: Three
color image of this IRDC: 8 $\mu$m emission in red, 4.5 $\mu$m
emission in green and 3.6 $\mu$m emission in blue. The two EGOs
within this IRDC are marked.}
\end{figure}

\section{DATA}
The Millimeter Astronomy Legacy Team 90 GHz survey is a large
international project aimed at characterizing the sites within our
Galaxy where high-mass stars will form. The survey covers a Galactic
longitude range  of $\sim$ -60 to $\sim$ 15$^{\circ}$ and Galactic
latitude range of -1 to +1 $^{\circ}$. The observations were carried
out with the newly upgraded Mopra Spectrometer (MOPS). The full 8
GHz bandwidth of MOPS was split into 16 zoom bands of 138 MHz each
providing a velocity resolution of $\sim$ 0.11 km s$^{-1}$. The
angular resolution of Mopra is about 38 arcsec, with beam efficiency
between 0.49 at 86 GHz and 0.42 at 115 GHz (Ladd et al. 2005). The
central frequencies of transitions we selected to study in the paper
are shown in Table 1. More information about this survey can be
found through the MALT90 website (http://malt90.bu.edu). The data
processing was conducted using Gildas and line parameters (peak
intensity, central velocity, FWHM) are obtained by Gaussian fitting.

To complement the molecular data we used data from Galactic Legacy
Infrared Mid-Plane Survey Extraordinaire (GLIMPSE) survey of Spitzer
to study young stellar object (YSO) in the region.

\begin{table}[htbp]
 \begin{center}
 \caption{\label{tab:test}List of the transitions selected to study in
this paper}
 \begin{tabular}{lclcl}
  \hline
  Species & Transition & $\nu$ (GHz)  &Primary Information Provided
\\
  \hline
  HCO$^+$ & (1-0) & 89.189 & High column density, kinematics\\
  H$^{13}$CO$^+$ & (1-0) & 86.754 & High column density, optical depth\\
  HNC & (1-0) & 90.663 & High column density, cold gas\\
  HC$_3$N & (10-9) & 91.200 & Hot core\\
  HNCO & (4$_{0,4}$-3$_{0,3}$) & 87.925 & Hot core\\
  C$_2$H & (1-0) & 87.317 & Photodissociation region\\
  SiO & (2-1) & 86.847 & Shock/outflow \\
  \hline
 \end{tabular}
 \end{center}
\end{table}

\section{ results and discussions}
Figure 1 (left) shows MSX A band image of MSXDC G331.71+00.59. The
locations of two EGOs are displayed on the right composite image of
Spitzer: 3.6 $\mu$m in blue, 4.5 $\mu$m in green, 8.0 $\mu$m in red.
G0.60 shows a larger size than G0.58.

\subsection{ molecular lines}
Figure 2 shows the molecular line spectra obtained towards G0.60 and
G0.58 respectively. It is worth noting that a single-gaussian fit is
insufficient for the optically thick molecular lines of HCO$^+$ and
HNC in the two sources that show prominent blue or red asymmetric
structures. Blue asymmetric structure, named ``blue profile'', a
combination of a double peak with a brighter blue peak or a skewed
single blue peak in optically thick lines (Mardones et al. 1997)has
been found in HCO$^+$ lines. This blue line asymmetry in an
optically thick tracer such as HCO$^+$ is often suggestive of
infall(i.e., Sun et al. 2008). Surely blue profile may also be
caused by rotation and outflow. However, infall motion is the only
process that would produce consistently the blue profile. Outflow
and rotation only produce a blue asymmetric line profile along a
particular line of sight to a source (Sun et al. 2008). Figure 4
shows mapping observation of HCO$^+$ of G0.58. The mapping
observation allows us to believe inflow activities in this region.
G0.60 has the similar image and is omitted to be displayed here.
Detections of optically thin line of H$^{13}$CO$^+$ help us to
determine the central velocity referred to as the Local Standard of
Rest, V$_{LSR}$. The parameters determined from Gaussian fitting of
these lines are presented in table 2. To quantify the blue profile,
we further use an asymmetry parameter $\delta$V defined as the
difference between the peak velocity of an optically thick line
V(thick) and an optically thin line V(thin) in units of the
optically thin line FWHM (Full Width at Half Maximum) dV(thin):
$\delta$V = $\frac{V(thick) - V(thin)}{dV(thin)}$. Mardones et al.
(1997) adopted a criterion $\delta$V $<$ -0.25 to indicate blue
profile. Our calculations demonstrate blue profile caused by inflow
in this IRDC.

\begin{figure}
\onecolumn \centering
\includegraphics[height=4.5in,width=3in,angle=270]{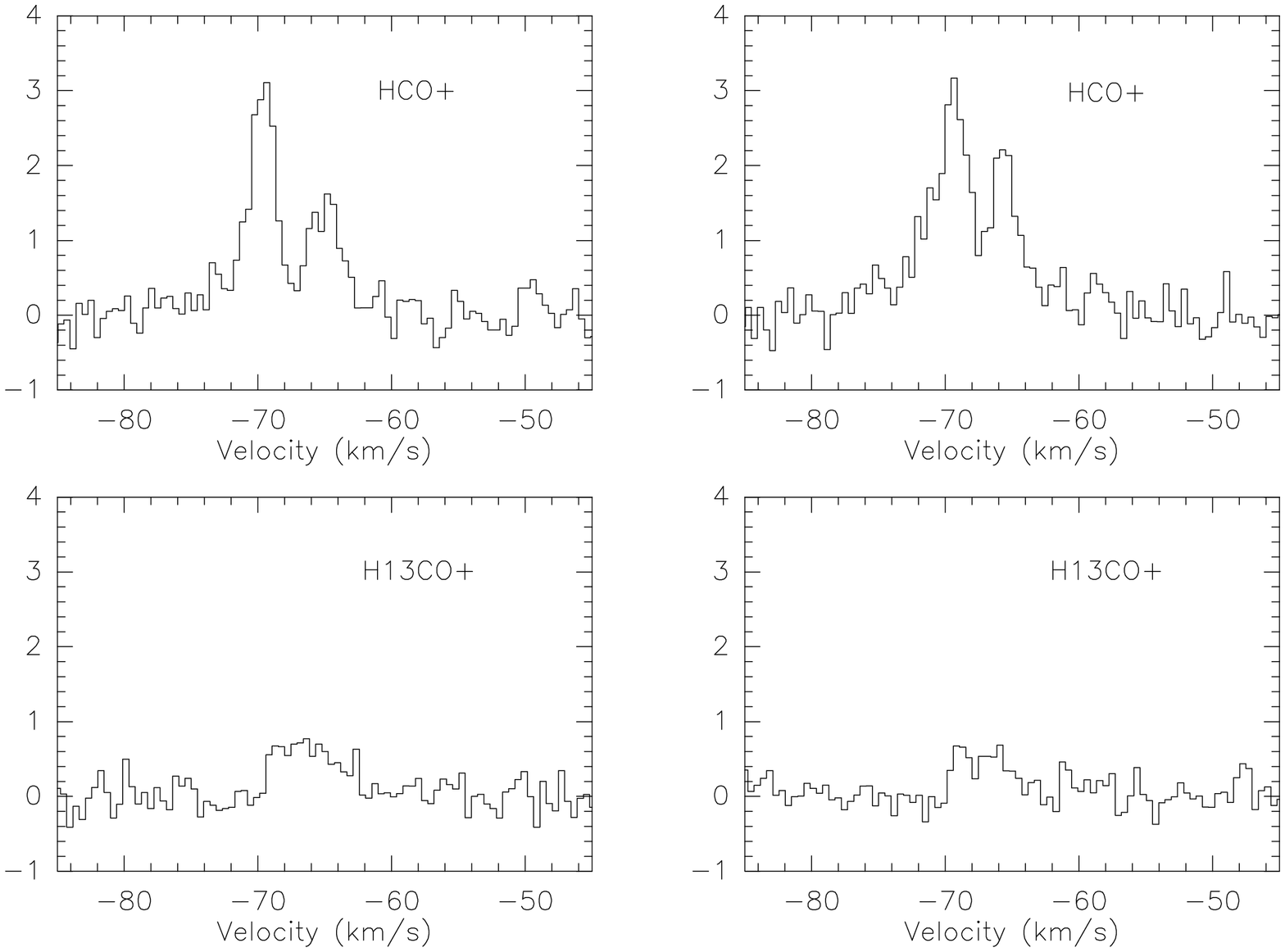}
\includegraphics[height=4.5in,width=3in,angle=270]{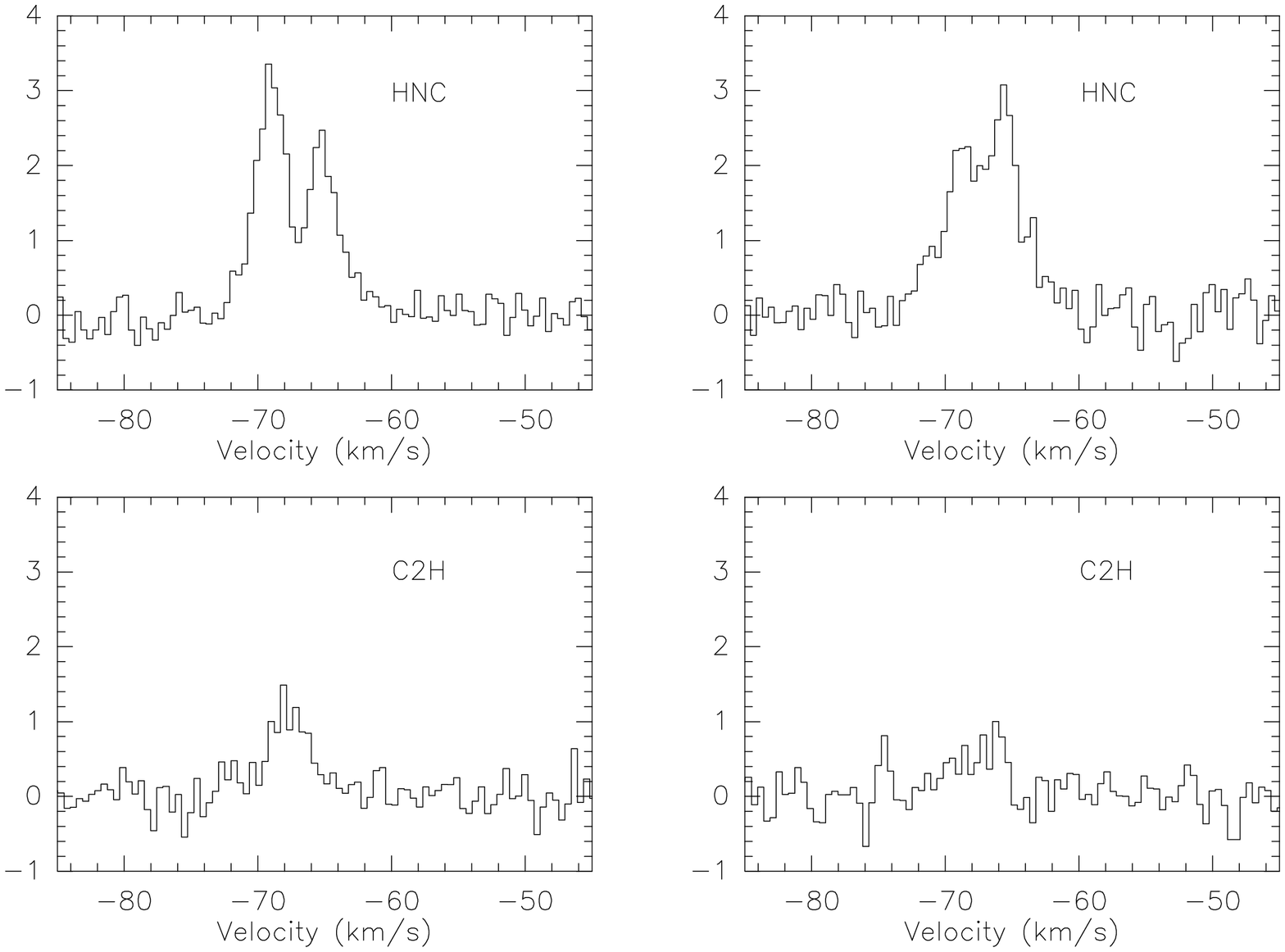}
\includegraphics[height=4.5in,width=3in,angle=270]{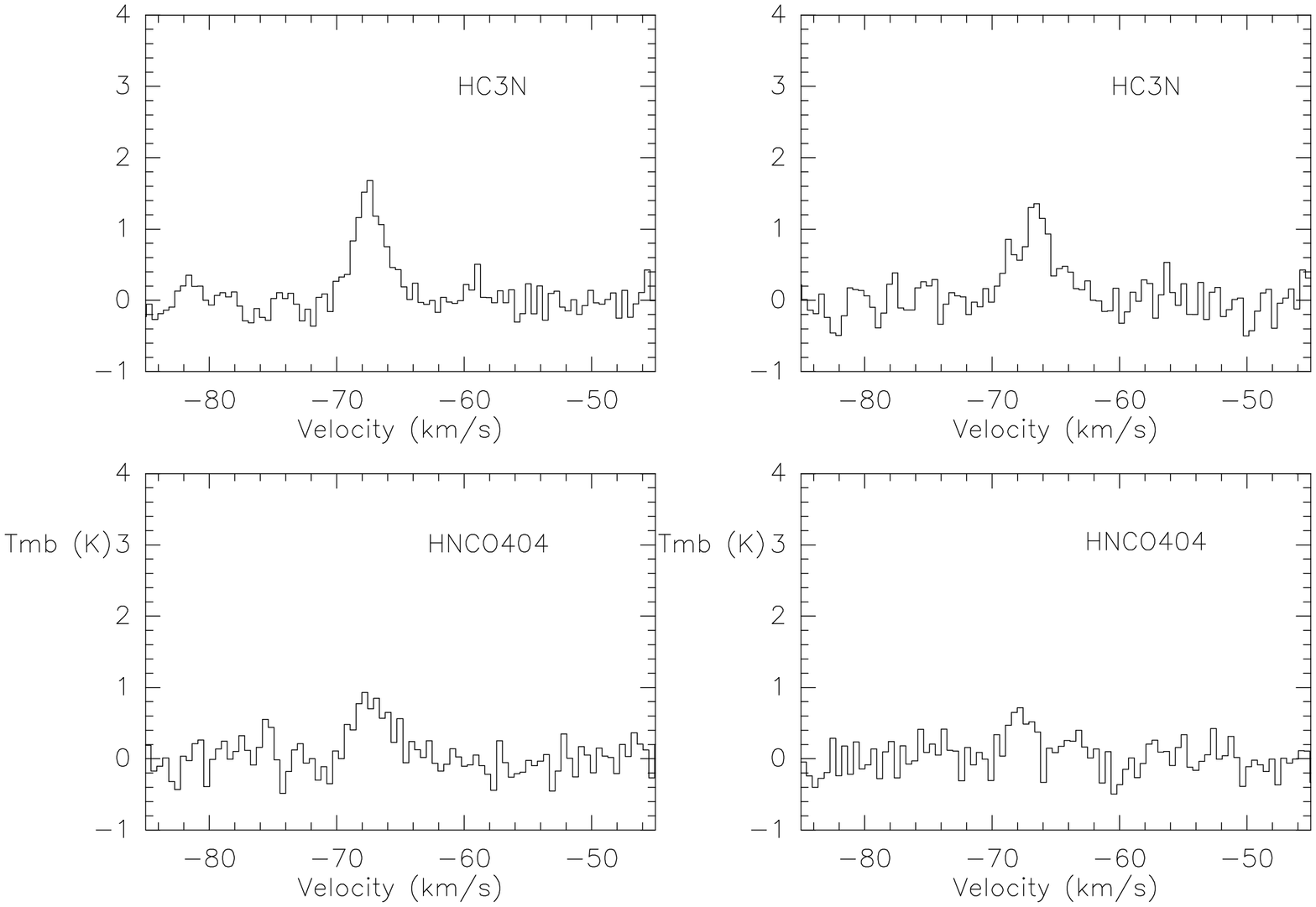}
\caption{Spectra towards the center of G0.60 (on the left) and G0.58
(on the right). The vertical axis is the main beam brightness
temperature.}
\end{figure}

\begin{figure}
\onecolumn \centering
\includegraphics[height=4in,width=7.5in]{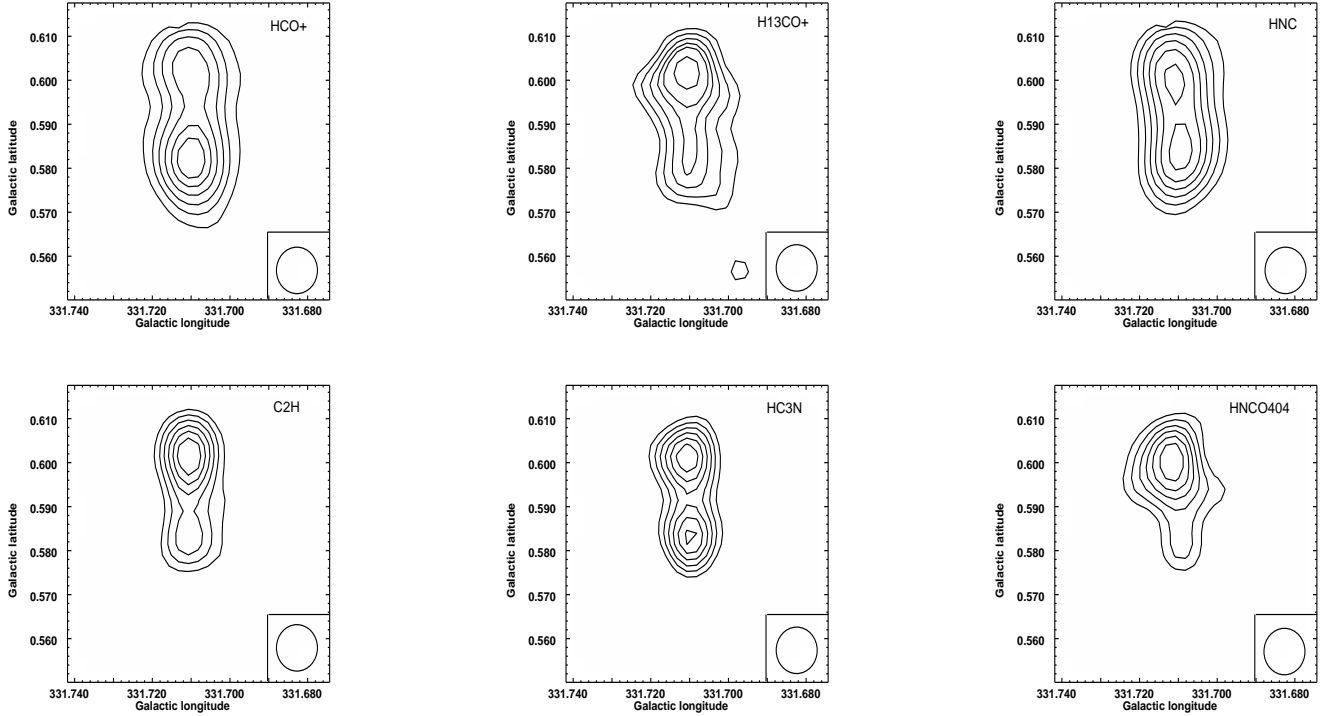}
\caption{Maps of the integrated intensity of the six transitions.
The contour levels are all 30, 40...90 $\%$ of each peak. The beam
of the Mopra telescope at the observation frequency is shown in the
right corner. }
\end{figure}

\begin{figure}
\onecolumn \centering
\includegraphics[height=4in,width=3.5in,angle=270]{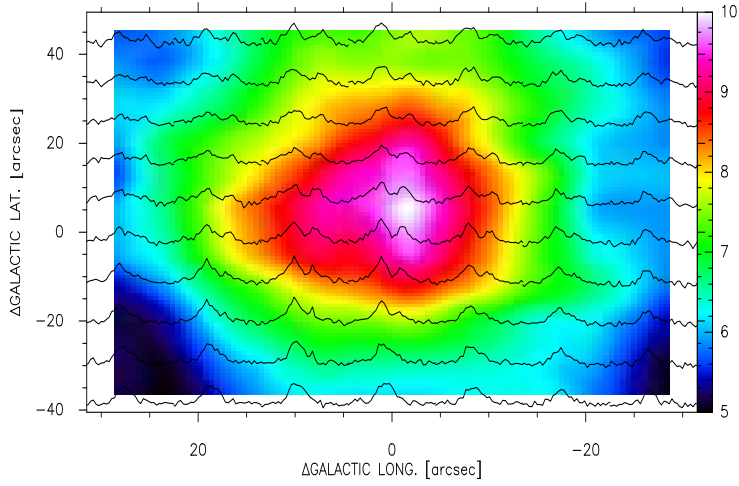}
\caption{HCO$^+$ spectra superimposed on its integrated map obtained
towards G0.58.}
\end{figure}

\begin{table}[htbp]
 \begin{center}
 \caption{\label{tab:test}Observed parameters of the molecular lines shown in Figure 2}
 \begin{tabular}{lclclclcl}
  \hline
  Source & Emission & T$_{mb}$ (K)  &V$_{LSR}$ (km s$^{-1}$) & FWHM
  (km s$^{-1}$)
\\
  \hline
  G0.60 & HCO$^+$ & 3.07 $\pm$ 0.08 & -69.66 $\pm$ 0.06 & 2.38 $\pm$ 0.16\\
        &         &1.52 $\pm$ 0.05 & -64.96 $\pm$ 0.13 & 2.88 $\pm$ 0.31\\
        & H$^{13}$CO$^+$ & 0.97 $\pm$ 0.12 & -66.32 $\pm$ 0.30 & 5.30 $\pm$ 0.61\\
        &  HNC  & 3.18 $\pm$ 0.06 & -69.06 $\pm$ 0.05 & 2.66 $\pm$ 0.13\\
        &     & 2.23 $\pm$ 0.09 & -65.10 $\pm$ 0.07 & 2.70 $\pm$ 0.22\\
        & C$_2$H & 1.15 $\pm$ 0.12 & -67.71 $\pm$ 0.17 & 3.77 $\pm$ 0.52 \\
        & HC$_3$N & 1.51 $\pm$ 0.08 & -67.43 $\pm$ 0.09 & 2.80 $\pm$ 0.22\\
        & HNCO & 0.86 $\pm$ 0.05 & -67.23 $\pm$ 0.22 & 3.22 $\pm$ 0.47\\
  \hline
  G0.58 & HCO$^+$ & 2.62 $\pm$ 0.40 & -69.56 $\pm$ 0.10 & 3.87 $\pm$ 0.34\\
        &         &2.06 $\pm$ 0.20 & -65.41 $\pm$ 0.11 & 2.14 $\pm$ 0.26\\
        & H$^{13}$CO$^+$ & 0.59 $\pm$ 0.03 & -67.08 $\pm$ 0.37 & 4.36 $\pm$ 0.71\\
        & HNC & 1.93 $\pm$ 0.20 & -68.81 $\pm$ 0.36 & 4.05 $\pm$ 0.77\\
        & & 2.30 $\pm$ 0.25 & -65.38 $\pm$ 0.21 & 3.00 $\pm$ 0.43\\
        &C$_2$H & 0.65 $\pm$ 0.10 & -67.50 $\pm$ 0.42 & 4.49 $\pm$ 0.91\\
        &HC$_3$N & 1.12 $\pm$ 0.15 & -66.67 $\pm$ 0.18 & 38.2 $\pm$ 0.46\\
        & HNCO & 0.71 $\pm$ 0.08 & -67.80 $\pm$ 0.20 & 2.04 $\pm$ 0.41\\
  \hline
 \end{tabular}
 \end{center}
\end{table}

\subsubsection{Column densities and masses}
We derive the molecular column densities of HCO$^+$ and hence masses
in the region assuming optically thick HCO$^+$ and optically thin
H$^{13}$CO$^+$ lines and local thermodynamic equilibrium (LTE). We
calculate the excitation temperature from
\begin{equation}
T_{ex} = \frac{hv_0}{k}[ ln(1 + \frac{hv_0/k}{T_{max}(HCO^+) +
J_v(T_{bg})})]^{-1}
\end{equation}
where $\nu$$_0$ is the rest frequency, T$_{bg}$ is temperature of
the background radiation (2.73 K) and
\begin{equation}
J_{v}(T) = \frac{hv_0}{k}\frac{1}{(e^{hv_0/kT} - 1)}
\end{equation}
 We also  assume the HCO$^+$ and H$^{13}$CO$^+$ emission arises from the same gas and
shares a common excitation temperature. The optical depth of the
H$^{13}$CO$^+$ line may be found from
\begin{equation}
\tau = - ln [1 - \frac{T_{max}(H^{13}CO^+)}{[J_v(T_{ex}) -
J_v(T_{bg})]}]
\end{equation}
Then we use equations (2) and (5) in Purcell et al. (2006) to find
the total H$^{13}$CO$^+$ column densities. Using M = $\mu$ \itshape
m$_H$ d$^2$ \upshape $\Omega$ X(H$^{13}$CO$^+$)$^{-1}$
N(H$^{13}$CO$^+$), we obtain the masses for the two cores, where
N(H$^{13}$CO$^+$) is the H$^{13}$CO$^+$ column density, \itshape d
\upshape the distance, \itshape m$_{H}$ \upshape the hydrogen atom
mass. We adopt a mean molecular weight per H$_2$ molecule of $\mu$ =
2.72 to include helium, $\Omega$ is the area of the two cores. For
H$^{13}$CO$^+$, we adopt an abundance of 2 $\times$ 10$^{-10}$
relative to H$_2$ (Vogel et al. 1984; Rawlings et al. 2004; Purcell
et al. 2006; Klaasen $\&$ Wilson 2007). Using the above methods we
obtain for T$_{ex}$ 5.9 K and 5.6 K respectively. These values are
considerably lower than the $\sim$15 K temperatures quoted in
previous work and we conclude that the emission is beam diluted in a
significant fraction of the observations. In the following analysis
we have assumed an excitation temperature of 15 K for HCO$^{+}$. The
final results are listed in table 3. The high column densities
suggest that massive stars are forming. Recent theoretical work
predicts a mass column density threshold $>$ 1 g cm$^{-2}$ for
massive star formation (Krumholz $\&$ McKee 2008; Krumholez et al.
2010).

\begin{table}[htbp]
 \begin{center}
 \caption{\label{tab:test}Derived parameters of G0.60 and G0.58}
 \begin{tabular}{lclclcl}
  \hline
  Source & T$_{ex}$ (HCO$^+$) (K) & $\tau$ (H$^{13}$CO$^+$) & N (H$^{13}$CO$^+$) ($\times$ 10$^{13}$ cm$^{-2}$) & M ($\times$ 10$^{3}$M$_\odot$)
\\
  \hline
G0.60 & 5.9 & 0.08 & 3.6 & 1.9\\
  \hline
G0.58 & 5.6 & 0.05 & 2.3 & 0.9\\
  \hline
 \end{tabular}
 \end{center}
\end{table}

\subsubsection{outflow}
It should be noted that the central HNC spectra of G0.58 shows a
``red profile''. The origin of red profile could be caused by
outflows or rotations of molecular gas. As EGOs are probably massive
young stellar objects (MYSOs) driving outflows, outflow activities
could exist in this region. Fig 5 presents the HNC (solid line),
H$^{13}$CO$^+$ (dashed line) spectra towards the center of the G0.58
and G0.60. The line wings of HNC are obvious. The existence of
outflow is further confirmed by position-velocity (PV) diagrams
shown in figure 5.
\begin{figure}
\onecolumn \centering
\includegraphics[height=2.5in,width=2in,angle=270]{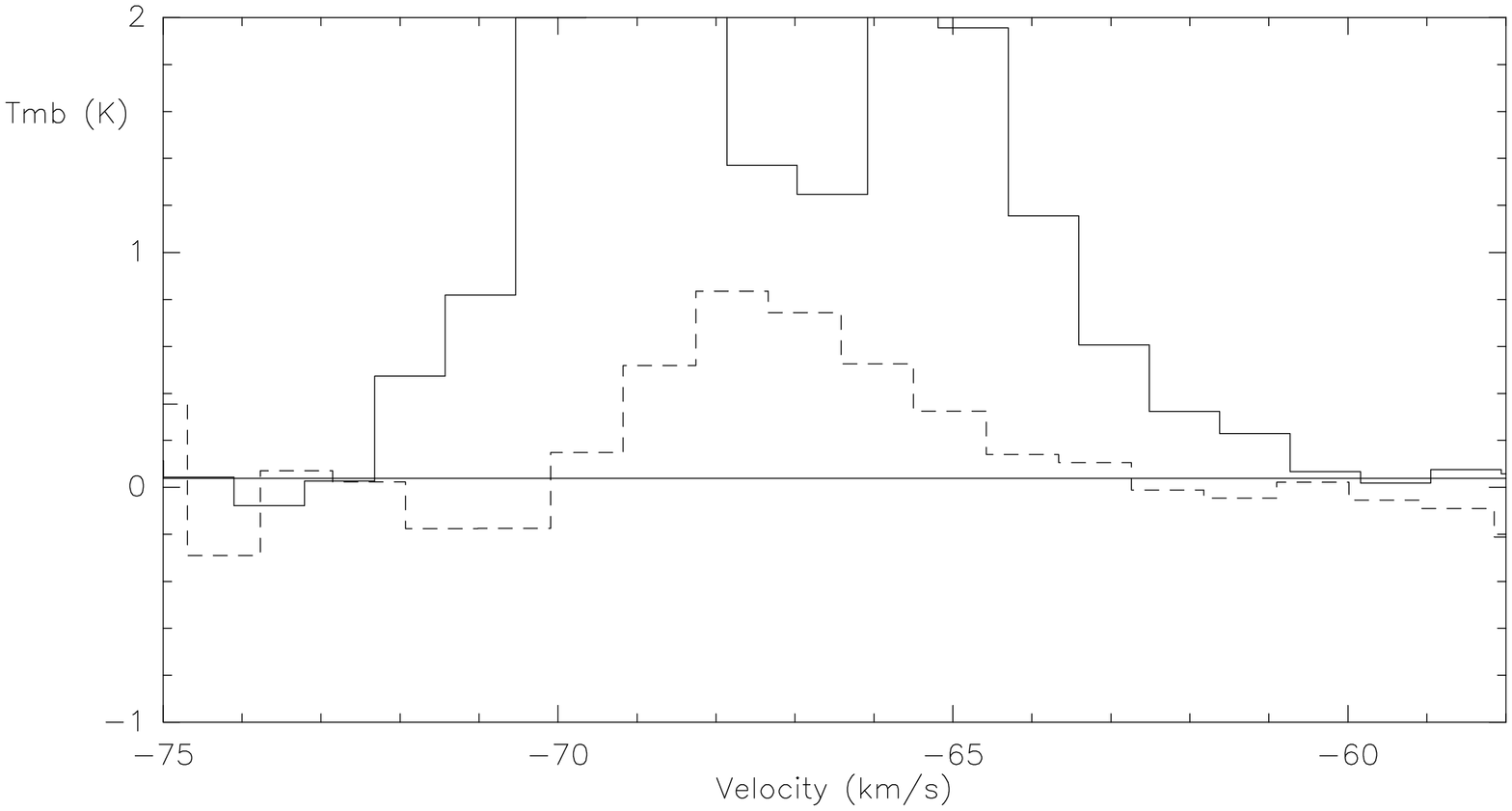}
\includegraphics[height=2.5in,width=2in,angle=270]{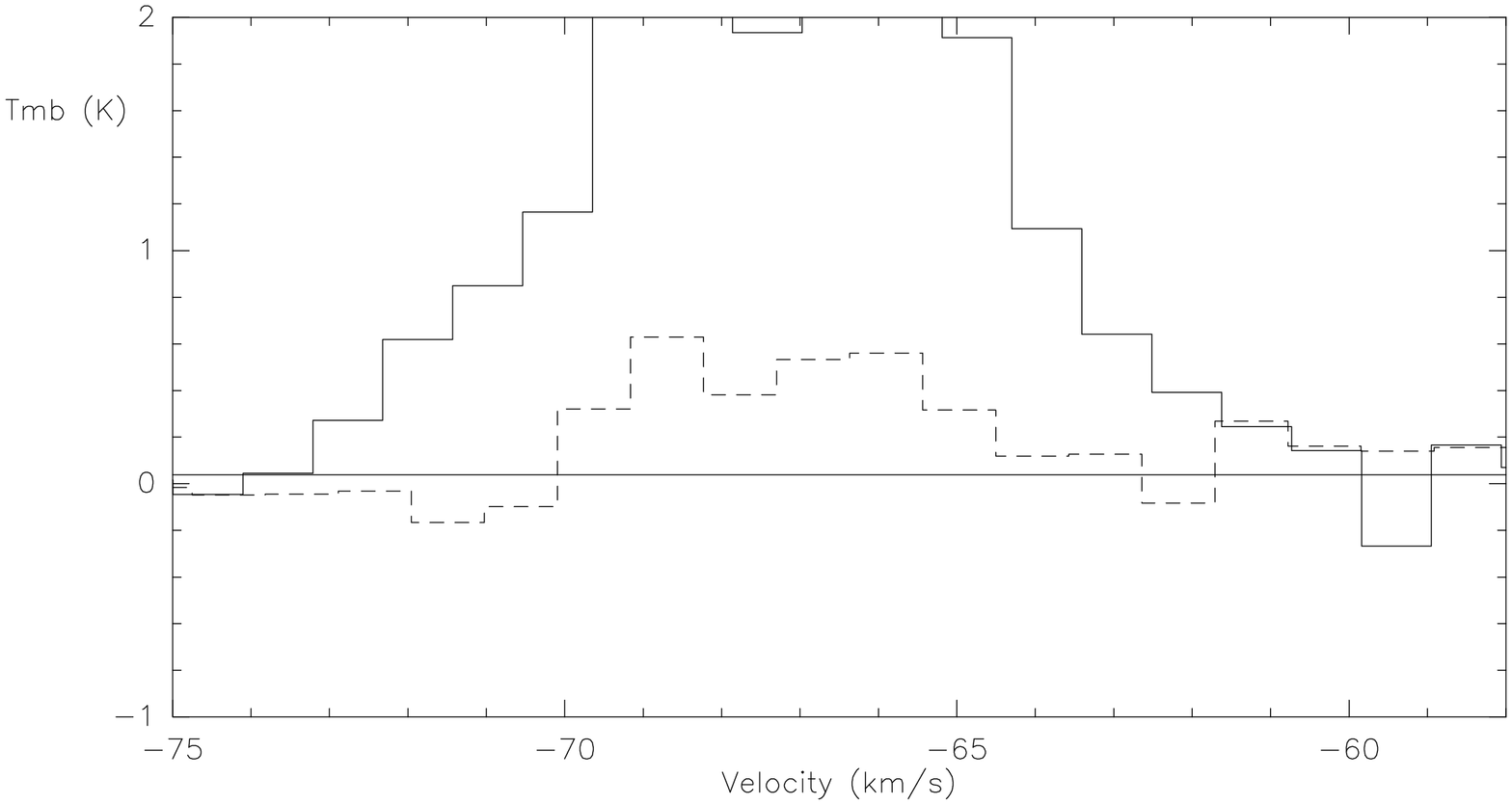}
\includegraphics[height=2.5in,width=2.5in,angle=270]{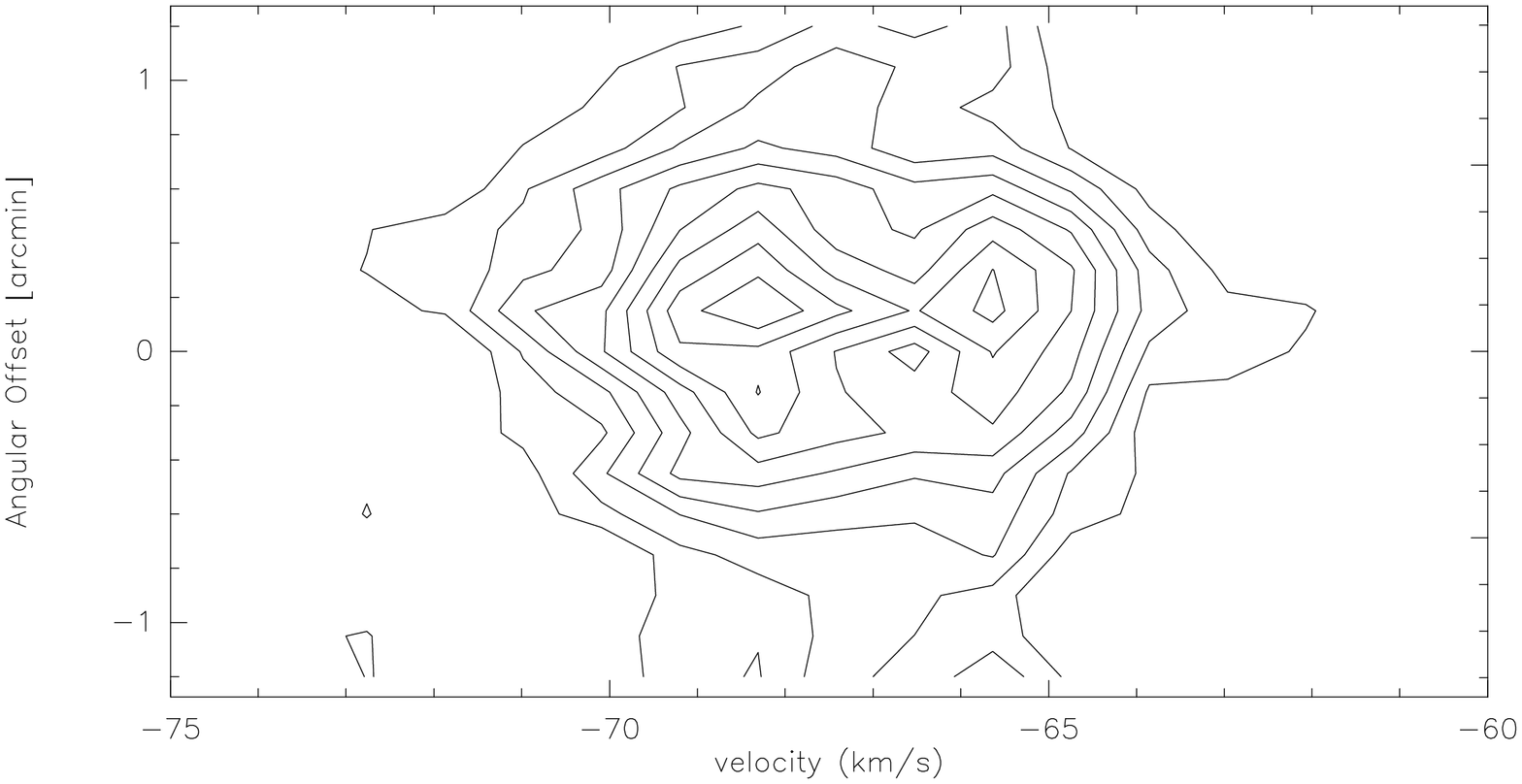}
\includegraphics[height=2.5in,width=2.5in,angle=270]{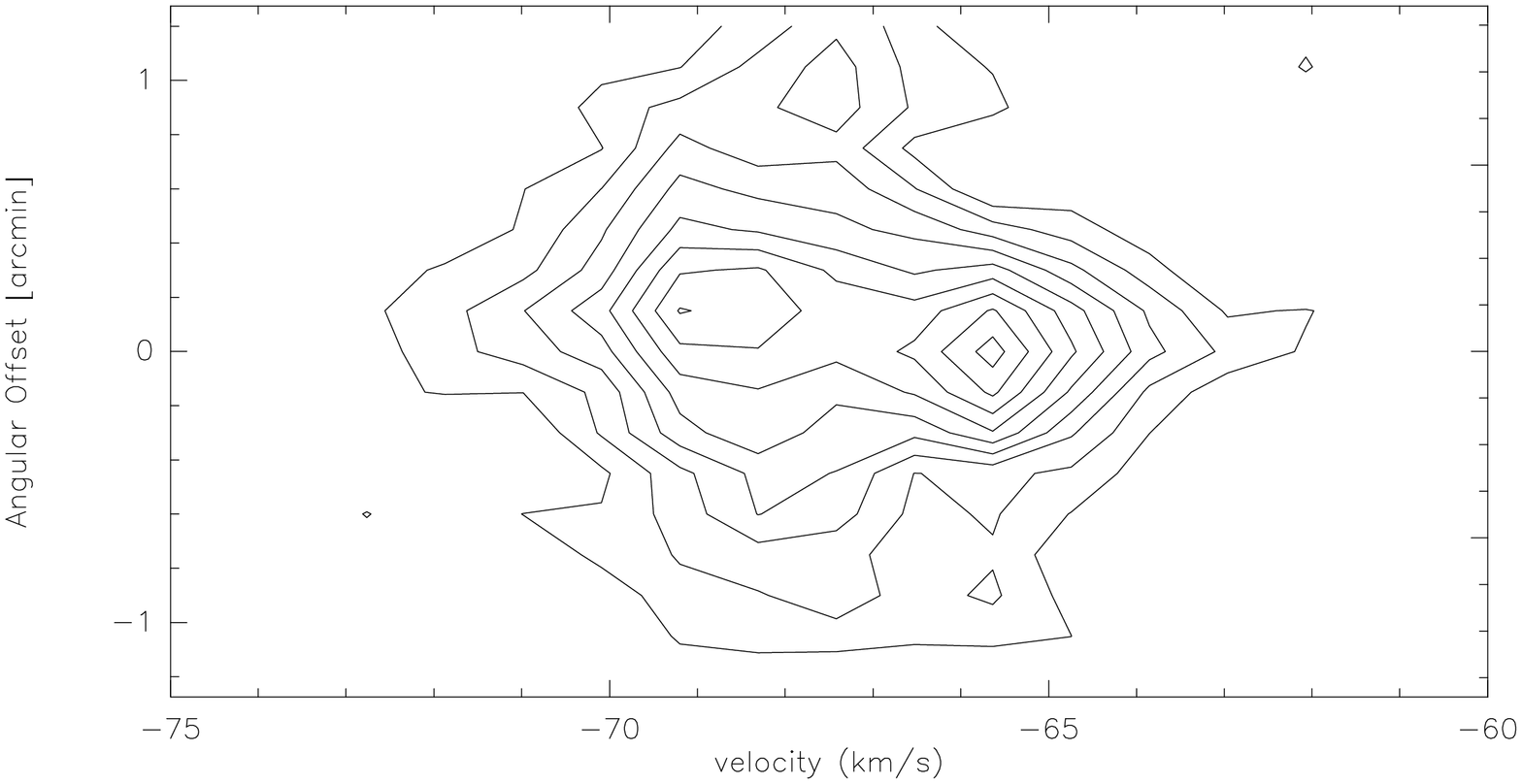}
\caption{Top: zoom-in of the HNC (solid line), H$^{13}$CO$^+$
(dashed line) profiles presented in Fig.2 in the intensity range
going from -1 to 2 K. Bottom: Position-velocity diagram cut along
east-west direction (shown in Fig. 6) constructed from HNC
transition of G0.60 and G0.58. Contour levels are 20, 30...90 $\%$
of the center. }
\end{figure}

\begin{figure}
\onecolumn \centering
\includegraphics[height=2.5in,width=3in]{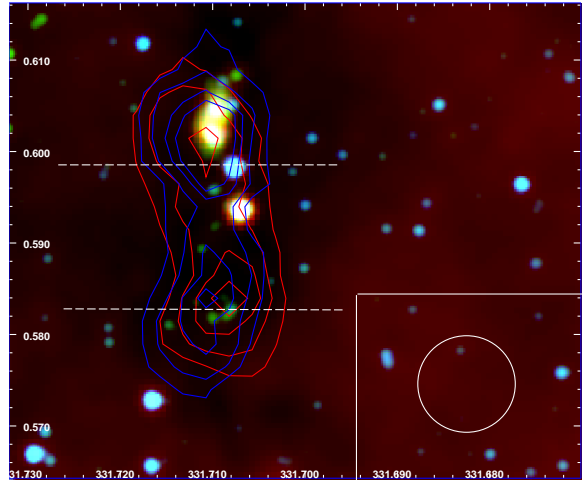}
\caption{Spitzer image of the IRDC G331.75+00.59 region. The blue
and red contours represent the integrated intensity of the HNC
transition in the blue and red shifted wings respectively. Contour
levels are 3.6, 4.6, 5.6, 6.6 K kms$^{-1}$. The two white dashed
lines are the position and orientation where the PV diagrams shown
in Fig. 5 were computed.}
\end{figure}

Figure 6 shows the map of the molecular outflow traced by the HNC
line; plotted are the intensities integrated over the velocity
ranges that are lower and high than the systematic velocity, i.e.,
red and blue shifted emission respectively. The red and blue
contours represent blue and red lobes of the outflow. From the HNC
map it can be noted that the lobes are nearly unresolved by our
observations, and that high resolution is needed to determine
parameters such as the collimation factor and inclination angles of
the outflows. Assuming that HNC emission in the line wings to be
optically thin, X(HNV) = [HNC]/[H$_2$] = 10$^{-8}$ (Turner et al.
1997) and T$_{ex}$ = \itshape hv$_0$/k \upshape =4.36 K, we derive
its column density from:
\begin{equation}
N(HNC) = Q(T_{ex}) \frac{8 \pi v_0^3}{c^3} \frac{g_l}{g_u}
\frac{1}{A_{ul}} [1 - e^{-hv_0/kT_{ex}}]^{-1} \int \tau dv
\end{equation}
where $\nu$$_0$, \itshape g$_u$, g$_l$\upshape and A$_{ul}$ are the
rest frequency, the upper and lower level degeneracies and the
Einstein's coefficient of HNC, Q(T$_{ex}$) is the partition
function, and \itshape c \upshape is the speed of light. On the
other hand, by assuming HNC emission is optically thin in the line
wings, we use the approximation:
\begin{equation}
\int \tau dv = \frac{1}{J(T_{ex}) - J(T_{bg})} \int T_{mb} dv
\end{equation}
The derived parameters are shown in table 4.

\begin{table}[htbp]
 \begin{center}
 \caption{\label{tab:test}Outflow parameters}
 \begin{tabular}{lclclcl}
  \hline
 Source & Shift & $\Delta$$\upsilonup$ (Km s$^{-1}$)& N (HNC) ($\times$ 10$^{13}$ cm$^{-2}$) & M (M$_{\odot}$)
\\
  \hline
G0.60 & red &(-66,-60)& 1.5 & 15.8\\
      & blue &(-75,-69)& 1.3& 13.7\\
  \hline
G0.58 & red &(-66,-60)&1.9 & 20.0\\
      & blue &(-75,-69)& 1.1& 11.6\\
  \hline
 \end{tabular}
 \end{center}
\end{table}

\subsubsection{other lines and SiO}
The ethynyl radical (C$_2$H) was first detected in the interstellar
clouds by Tucker et al. (1974). Observations indicate that C$_2$H is
produced in photodissociation regions (e.g., Lo et al. 2009; Gerin
et al. 2011) and almost omnipresent toward evolutionary stages from
infrared dark clouds (IRDC) via high-mass protostellar objects
(HMPO) to ultracompact HII regions (UCHII) (Huggins et al. 1984;
Beuther et al. 2008). The stronger emission of C$_2$H in G0.60 than
G0.58 (figure 3) suggests G0.58 to be in an earlier evolutionary
stage. This was further confirmed by carbon-bearing species HNCO,
which is typically only seen in the hot cores around high-mass
protostars once molecule has been liberated off dust grains by
radiation or shocks (Brown et al. 1988). HC$_3$N is an excellent
dense gas tracer (Chung et al. 1991; Bergin et al. 1996) as it has
high electric dipole moments ($\mu$ = 3.72 Debye). It is also a
tracer of hot core because it can be easily destroyed by ultraviolet
(UV) photons from central ionizing stars (Brown et al. 1988). Thus
we regard the two EGOs are evolving from IRDC to hot cores.

SiO is known to be greatly enhanced in outflows and shocked regions
(Martin - Pintado, Bachiller $\&$ Fuente 1992) since in the general
interstellar medium Si is frozen out onto dust grains. When the gas
in a region is shocked, for example the gas through which a
protostellar outflow is passing or due to the expanding of
photodissociation region (PDR), the dust grains can sublimate and Si
is released into the gas phase. Models and observations of SiO in
PDRs around high massive star formation regions suggest moderate SiO
enhancement. Schilke et al. (2001) find SiO column densities of
$\sim$ 10$^{12}$ cm$^{-2}$ in their observed PDRs. Again, by
assuming SiO is optically thin and LTE, we calculated the column
density as the method described in section 3.1.2. The high SiO
column densities (12.2 $\times$ 10$^{14}$ cm$^{-2}$ for G0.60 and
18.6 $\times$ 10$^{14}$ cm$^{-2}$ for G0.58 ) suggest that the SiO
emission we observed does come primarily from outflow shocks. The
large line width of SiO ($>$ 10 km s$^{-1}$) further supports our
suggestions. Figure 7 shows the integrated intensity map of SiO and
spectra towards the center of G0.60 and G0.58.

\begin{figure}
\onecolumn \centering
\includegraphics[height=3.in,width=2.3in,angle=-90]{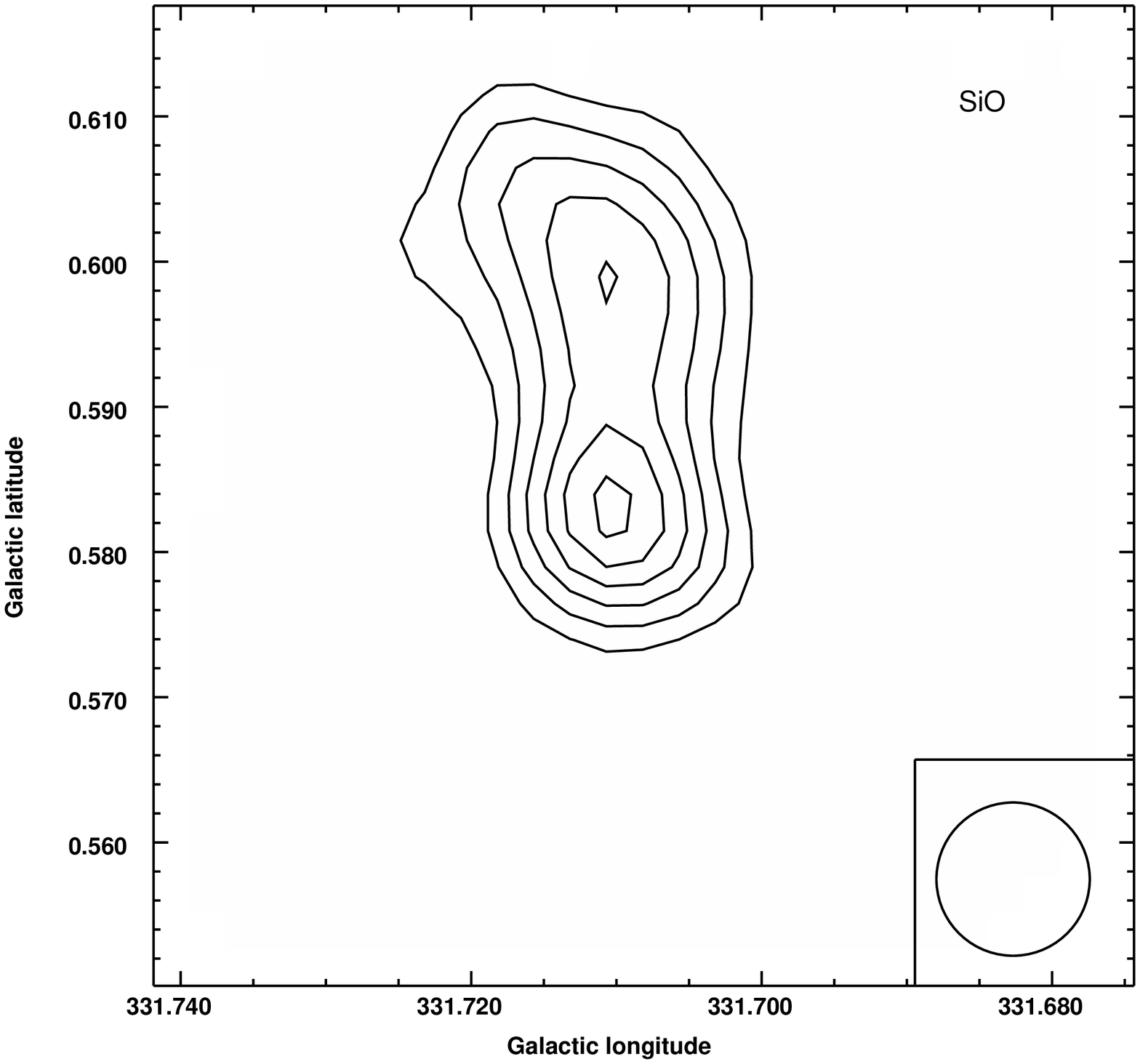}
\includegraphics[height=2.65in,width=2.8in,angle=-90]{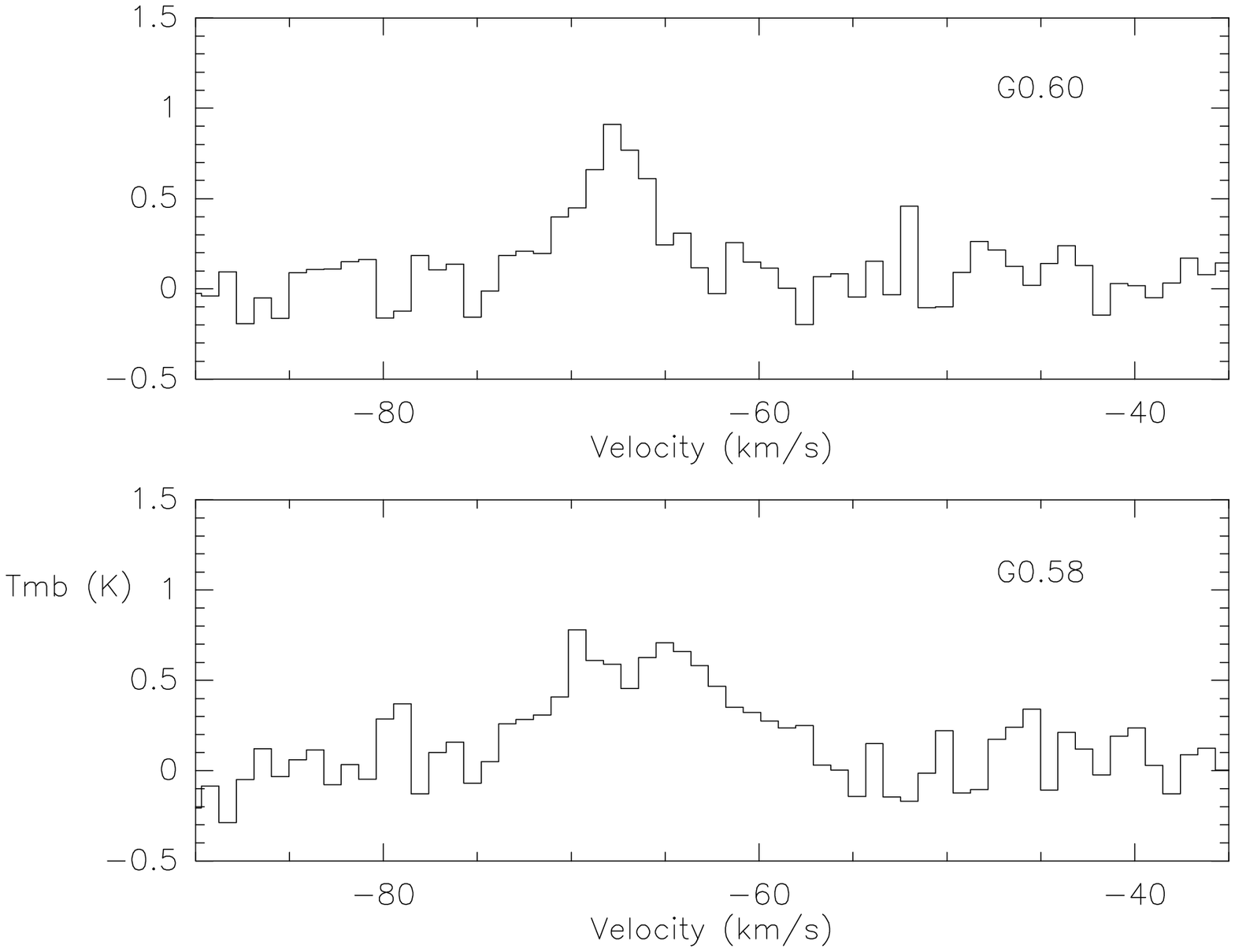}
\caption{Left: Integrated intensity map of SiO. Contour levels are
30, 40...90 $\%$ of the peak. Right: SiO spectra towards the center
of G0.60 and G0.59}
\end{figure}

\subsection{spectral energy distribution}
According to Cyganowski et al. (2008), the mid-IR colors of EGOs lie
in regions of color-color diagram (CCD) occupied by young protostars
still embedded in infalling envelopes. In this section, we try to
fit the spectral energy distribution (SED) of these two EGOs using
the tool developed by Robitaille et al. (2007). Briefly, the
SED-fitting tool works as a regression method to find the SEDs
within a specified $\chi$$^{2}$ from a large grid of models after
fitting the input data points. The grid of models contains stellar
masses, disk masses, mass accretion rates, and line-of-sight (LOS)
inclinations. The grid of YSO models was computed by Robitaille et
al. (2006) using the 20,000 two-dimensional radiation transfer
models from Whitney et al. (2003a, 2003b, 2004). Each YSO model has
SEDs for 10 viewing angles (inclinations), so the total YSO grid
consists of 200,000 SEDs. We use the fluxes in the four Spitzer IRAC
bands and MIPS in 24 $\mu$m from Cyganowski et al. (2008). The 4.5
$\mu$m band is used as an upper limit, considering it is blended
with emissions from outflows. The up left panel in figure 8 shows
the SEDs of the 20 best fitting models of G0.60. The solid black
line represents the best fitting model from which we obtain a
central mass of $\sim$ 10 M$_\odot$. The SED fitting result of G0.58
is not good enough, due to large errors in 5.8 $\mu$m and 8.0 $\mu$m
(Cyganowski et al. 2008). Figure 8 also presents the histograms with
the distribution of the constrained physical parameters. The hashed
columns represent the 20 best fitting models and the gray columns
correspond to all the models of the grid. Although there is a high
dispersion in the age, the presence of a massive envelope is a
strong evidence that the source is at an early evolutionary stage.
We note that MYSO(s) is forming in G0.60.

\begin{figure}
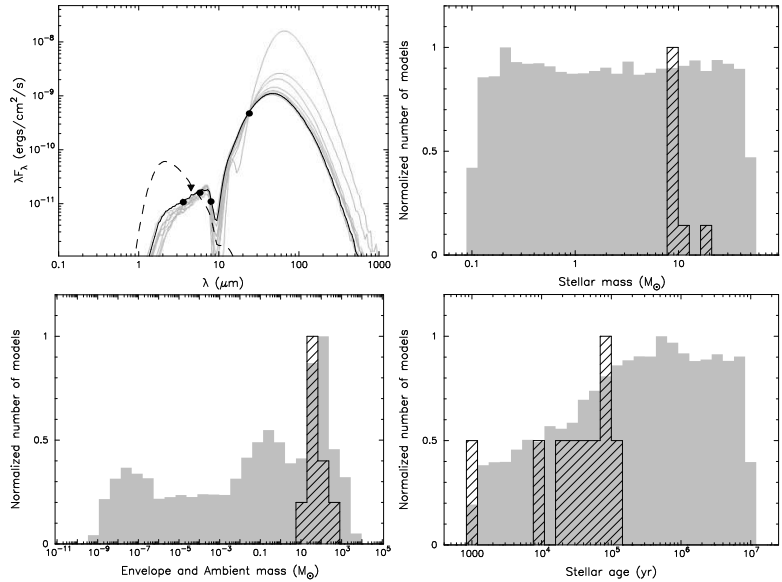

\onecolumn \centering
\includegraphics[height=1.5in,width=2in]{SED.eps}
\includegraphics[height=1.5in,width=2in]{MASS.eps}
\includegraphics[height=1.5in,width=2in]{ENVELOPE.eps}
\includegraphics[height=1.5in,width=2in]{age.eps}
\caption{Spectral energy distributions of G0.60 and distributions of
some physical parameters (central source mass, envelope mass and
age) of the best fitting models (hashed columns) together with the
distribution of all the models (grey columns). }
\end{figure}

\section{summary}
Using archive data from the MALT90, we made a first multi-transition
molecular line study of infrared dark cloud (IRDC) MSXDC
G331.71+00.59, where two EGOs embedded. Two molecular cores were
found to be associated with these EGOs. The HCO$^{+}$ (1-0) and HNC
(1-0) transitions show outflow and/or inflow activities of young
stellar objects. By the analysis of other molecules include
H$^{13}$CO$^{+}$ (1-0), C$_2$H (1-0), HC$_3$N (10-9),
HNCO(4$_{0,4}$-3$_{0,3}$) we regard the two EGOs are evolving from
IRDC to hot cores. Using public GLIMPS data, we investigate the
spectral energy distribution of EGO G331.71+0.60, supporting EGO is
a massive young stellar stellar object (MYSO) driving outflows.EGO
G331.71+0.58 may be in an earlier evolutionary stage. Moreover,
given its location in sky, G331.71+0.58 will easily be accessible in
the near future for high linear resolution studies with ALMA.

\section*{Acknowledgement}
We thank the MALT90 project team for the observations. We are also
grateful to the anonymous referee for whose constructive
suggestions.


\begin{thebibliography}{}
\bibitem[Blitz et al. (1982)]{b2} Bergin, E. A., Snell, R. L., Goldsmith, P. F. 1996, ApJ, 460, 343
\bibitem[Blitz et al. (1982)]{b2} Beuther, H., Semenov, D., Henning, T., Linz, H. 2008, ApJ, 675, L33
\bibitem[Blitz et al. (1982)]{b2} Brown, P. D., Charnley, S. B., Millar, T. J. 1988, MNRAS, 231, 409
\bibitem[Blitz et al. (1982)]{b2} Chung, H. S., Kameya, O., Morimoto, M. 1991, JKAS, 24, 217
\bibitem[Blitz et al. (1982)]{b2} Cyganowski, C. J., Whitney, B. A., Holden, E., et al. 2008, AJ, 136, 2391
\bibitem[Blitz et al. (1982)]{b2} Egan, M. P., Shipman, R. F., Price, S. D.,
Carey, S. J., Clark, F. O., Cohen, M., 1998, ApJ, 494, L199
\bibitem[Blitz et al. (1982)]{b2} Fich, M., Blitz, L., Stark, A. A., 1989, ApJ,
\bibitem[Blitz et al. (1982)]{b2} Gerin, M., Ka$\acute{z}$mierczak, M., Jastrzebska, M., et al. 2011,
A\&A, 525, A116+
\bibitem[Blitz et al. (1982)]{b2} Huggins, P. J., Carlson, W. J., Kinney, A. L.
1984, A\&A, 133, 347
\bibitem[Blitz et al. (1982)]{b2} Jackson, J.M., Finn, S.C., Rathborne, J.M.,
Chambers, E.T., Simon, R., 2008, ApJ, 680, 349
\bibitem[Blitz et al. (1982)]{b2} Klaassen, P.D., Wilson, C.D., 2007, ApJ, 663,
1092
\bibitem[Blitz et al. (1982)]{b2} Krumholz, M.R., Cunningham, A.J., Klein, R.I.,
McKee, C.F., 2010, ApJ, 713, 1120
\bibitem[Blitz et al. (1982)]{b2} Krumholz, M.R., McKee, C.F., 2008, Nature,
451, 1082
\bibitem[Blitz et al. (1982)]{b2} Ladd, N., Purcell, C., Wong, T., \& Robertson, S. 2005, PASA, 22, 62
\bibitem[Blitz et al. (1982)]{b2} Lo, N., Cunningham, M. R., Jones, P. A., et al. 2009, MNRAS, 395,
1021
\bibitem[Blitz et al. (1982)]{b2} Mardones, D., Myers, P. C., Tafalla, M. et al. 1997, ApJ, 489, 719
\bibitem[Blitz et al. (1982)]{b2} Martin - Pintado, J., Bachiller, R., Fuente,
A., 1992, A\&A, 254, 315
\bibitem[Blitz et al. (1982)]{b2} Perault, M., Omont, A., Simon, G. et al.,
1996, A\&A, 315, L165
\bibitem[Blitz et al. (1982)]{b2} Purcell, C. R., Balasubramanyam, R., Burton,
M. G., et al., 2006, MNRAS, 367, 553
\bibitem[Blitz et al. (1982)]{b2} Rawlings, J.M.C., Redman, M.P., Keto, E.,
Williams, D.A., 2004, MNRAS, 351, 1054
\bibitem[Blitz et al. (1982)]{b2} Reach, W.T., et al. 2006, AJ, 131, 1479
\bibitem[Blitz et al. (1982)]{b2} Robitaille, T. P., Whitney, B. A., Indebetouw, R., Wood,
K., 2007, ApJS, 169, 328
\bibitem[Blitz et al. (1982)]{b2} Robitaille, T. P., Whitney, B. A., Indebetouw, R., Wood, K.,  Denzmore, P., 2006, ApJS, 167, 256
\bibitem[Blitz et al. (1982)]{b2} Schilke, P., Pineau des For$\acute{e}$ts, G.,
Walmsley, C.M., et al., 2001, A\&A, 372, 291
\bibitem[Blitz et al. (1982)]{b2} Simon R., Jackson J.M., Rathborne J.M.,
Chambers E.T., 2006, ApJ, 639, 227
\bibitem[Blitz et al. (1982)]{b2} Sun Yan, Gao Yu, 2009, MNRAS, 392, 170
\bibitem[Blitz et al. (1982)]{b2} Tucker, K. D., Kutner, M. L., Thaddeus, P. 1974, ApJ, 193, L115
\bibitem[Blitz et al. (1982)]{b2} Turner, B.E., Pirogov, L., Minh, Y.C., 1997,
ApJ, 483, 235
\bibitem[Blitz et al. (1982)]{b2} Vogel, S.N., Wright, M.C.H., Plambeck, R.L.,
Welch, W.J., 1984, ApJ, 283, 655
\bibitem[Blitz et al. (1982)]{b2} Whitney, B. A., Indebetouw, R., Bjorkman, J.E., Wood, K., 2004, ApJ, 617, 1177
\bibitem[Blitz et al. (1982)]{b2} Whitney, B. A., Wood, K., Bjorkman, J. E., Cohen, M., 2003a, ApJ, 598, 1079
\bibitem[Blitz et al. (1982)]{b2} Whitney, B. A., Wood, K., Bjorkman, J. E., Wolff, M. J., 2003b, ApJ, 591, 1049


\end{thebibliography}
 \end{document}